\documentclass[aps,pra,reprint,groupedaddress]{revtex4-2}
\pdfoutput=1
\usepackage{amsmath}
\usepackage{amsfonts}
\usepackage{graphicx}
\usepackage{hyperref}

\usepackage{doi}

\newcommand{\bq}{q}
\newcommand{\bp}{p}

\newcommand{\bx}{x}

\newcommand{\hQ}{\mathcal{Q}}
\newcommand{\hC}{\mathcal{C}}
\newcommand{\hN}{\mathcal{N}}

\newcommand{\oI}{\Hat{I}}

\newcommand*{\defeq}{\stackrel{\text{def}}{=}}

\begin{document}

\title{Classical and semi-classical limits in phase space}

\author{Clay D. Spence} \email{clay.spence@sri.com}
\thanks{ORCiD: 0000-0002-8316-8786\\
  This paper was not part of any funded program at SRI International.}

\affiliation{SRI International\\
201 Washington Road\\
Princeton, NJ 08540}

\date{April 27, 2024}

\begin{abstract}
  A semiclassical approximation is derived by using a family of
  wavepackets to map arbitrary wavefunctions into phase space. If the
  Hamiltonian can be approximated as linear over each individual
  wavepacket, as often done when presenting Ehrenfest's theorem, the
  resulting approximation is a linear first-order partial differential
  equation on phase space, which will be referred to as the
  Schr\"odinger-Ehrenfest or SE equation. This advectively transports
  wavefunctions along classical trajectories, so that as a trajectory
  is followed in time the amplitude remains constant while the phase
  changes by the action divided by $\hbar$. The wavefunction's
  squared-magnitude is a plausible phase space density and obeys
  Liouville's equation for the classical time evolution. This is a
  derivation of the Koopman-von~Neumann (KvN) formulation of classical
  mechanics, which previously was postulated but not derived.  With
  the time-independent SE equation, continuity of the wavefunction
  requires the change of phase around any closed path in the torus
  covered by a classical trajectory to be an integer multiple of
  $2\pi$, giving a standing wave picture of old quantum
  mechanics. While this applies to any system, for separable systems
  it gives Bohr-Sommerfeld quantization.
\end{abstract}

\maketitle

\section{Introduction}

Semiclassical approximations are important not only for their
conceptual value in showing the relationship between classical and
quantum mechanics, but also for their practical applications
\cite{heller2018semiclassical}. A standard semiclassical approach is
to set $\psi = A \exp(\frac{i}{\hbar}S)$
\cite{keller58},\cite[Section~VI.4]{messiah1970quantum} or $\psi =
\exp(\frac{i}{\hbar}S)$ \cite[Section~7.5]{heller2018semiclassical},
insert into Schr\"odinger's equation, cancel the common exponential
factor and expand in powers of $\hbar$, considering only the
lowest-order terms. The results include classical flow in position
space, the Hamilton-Jacobi equation, and quantization.

In prior work \cite{torres1990quantum, harriman1994quantum,
  ban1998phase} that might appear to be unrelated, wavefunctions in
the usual Hilbert space are mapped into a larger Hilbert space
representable by wavefunctions on phase space. That prior work focuses
on developing quantum mechanics in terms of these phase-space
wavefunctions.

In this paper phase-space wavefunctions are used to provide another
semiclassical approximation. When Schr\"odinger's equation is mapped
into phase space, this approximation changes it to a first-order
partial differential equation \eqref{eq:classwave1}, which will be
referred to as the Schr\"odinger-Ehrenfest or SE equation. This
transports the wavefunction along classical trajectories, with the
amplitude remaining constant while the phase changes by the action
divided by $\hbar$. This is equivalent to the Koopman-von~Neumann
(KvN) formulation of classical mechanics
\cite{mauro2002koopman,koopman1931hamiltonian,neumann1932operatorenmethode},
except the phase plays no role in the latter.  The time-independent SE
equation gives quantization, since continuity of the wavefunction
restricts the phase to change by an integer multiple of $2\pi$ around
any closed path, whether it is a classical trajectory or not. While
that applies to any system, for separable systems the Bohr-Sommerfeld
quantization formula is derived. This lacks the zero-point
term. It will be shown that for the harmonic oscillator this term is
given by the second-order corrections in the SE approximation, so that
is the case for at least some systems.

\section{Approximating Schr\"odinger's equation on phase space}

In this section we give a fairly simple derivation of the first-order
Schr\"odinger-Ehrenfest equation. This implies Liouville's
equation. It is not quite the same as the KvN formulation of classical
mechanics, but we show that it can be changed into that form. We then
discuss the fact that the set of wavefunctions on phase space forms a
superspace of the conventional Hilbert space. We skipped this when first
deriving the SE equation for the sake of simplicity. The mapping
between conventional wavefunctions and wavefunctions on phase space is
mostly in the literature, and we review those sources. We end this
section with a presentation of second-order terms in the SE
approximation to Schr\"odinger's equation.

\subsection{The first-order SE equation\label{sec:cfromq}}

Consider a single particle in $D$ spatial dimensions with no internal
state like spin. In the notation used here when a variable with
multiple components is written without an index it indicates the
collection of all components. Schr\"{o}dinger's equation is
\( 
i\hbar\,\partial\psi(\bx, t) / \partial t = H(\bx, \Hat{\bp}, t)\psi(\bx,t).
\) 
Suppose that we measure position $\bx$ and momentum $\Hat{\bp}$
simultaneously, with some measurement errors.  If this measurement is
otherwise ideal it must project the state onto one with properties
consistent with the measurement results, so that the position and
momentum are concentrated around the observed values. Use $\bq$ for the
position coordinate in phase space to distinguish it from the argument
$\bx$ of the usual wavefunction. Note that up until
Section~\ref{sec:bohrsommerfeld} Cartesian coordinates are being used,
so through the sections before that $\bq$ are also Cartesian. Suppose
that after observing the position $\bq$ and momentum $\bp$ the resulting
state is the wave packet
\begin{equation}
\label{eq:udef}
u_{\bq\bp}(\bx) = u_0(\bx-\bq)
\exp\bigl(\tfrac{i}{\hbar}\bp \cdot (\bx-\bq)\bigr).
\end{equation}
Here $u_0$ is a wave packet with zero mean position and momentum and
variances $\sigma_x^2$ and $\sigma_p^2$, and the notation $a\cdot b =
\sum_i a_i b_i$ is used. The $u_{\bq\bp}$ have $\bq$ and $\bp$ for
means and the same variances as $u_0$.

The wavefunction can be expanded in a set of $u_{\bq\bp}$ as
\begin{equation}\label{eq:lift0}
\eta(\bq,\bp,t) = \int\!d^D\!x\, u^*_{\bq\bp}(\bx) \psi(\bx,t).
\end{equation}
We can show that the $u_{\bq\bp}$'s are self-inverting, i.e.,
\begin{equation}\label{eq:self-invert}
  \int \Bigl(\frac{dq\, dp}{2\pi\hbar}\Bigr)^D\,u_{\bq\bp}(\bx)u^*_{\bq\bp}(\bx')
  = \delta^D(\bx-\bx'),
\end{equation}
by integrating $\bp$ first to get the delta function. So this expansion
is invertible, and $\psi$ can be recovered from $\eta$ through
\begin{equation}\label{eq:project0}
\int \Bigl(\frac{dq\, dp}{2\pi\hbar}\Bigr)^D\, u_{\bq\bp}(\bx)\eta(\bq,\bp,t) = \psi(\bx,t).
\end{equation}

From Schr\"odinger's equation and the Hermiticity of $H$, the time
derivative of $\eta$ is
\begin{equation}
\frac{\partial \eta}{\partial t} =
 -\frac{i}{\hbar}\int\!dx\, [H(\bx, \Hat{\bp}, t)u_{\bq\bp}(\bx)]^* \psi(\bx,t).
\end{equation}
If $H$ varies slowly over the widths $\sigma_q$ and $\sigma_p$ of the
$u_{\bq\bp}$ it can be expanded in powers of $\bx-\bq$ and $\Hat{\bp}-\bp$. To
first order this gives
\begin{equation}
\begin{split}
\frac{  \partial \eta(\bq,\bp)}{\partial t} & = -\frac{i}{\hbar}
  \Bigl\{
  H(\bq,\bp,t)\eta(\bq,\bp) \\
  + & \sum_i \frac{\partial H}{\partial q_i}
  \int\!d^D\!x\, [(x_i -q_i)u_{\bq\bp}(\bx)]^* \psi(\bx,t) \\
  + & \sum_i \frac{\partial H}{\partial p_i}
  \int\!d^D\!x\, [(\Hat{p}_i-p_i)u_{\bq\bp}(\bx)]^* \psi(\bx,t)
  \Bigr\},
\end{split}
\label{eq:step1}
\end{equation}
where the sum is over all coordinate-momentum pairs, as usual. From
\eqref{eq:udef} it can be shown that
\begin{align}
  \label{eq:xu1}
  (x_i-q_i) u_{\bq\bp}(\bx) &= \frac{\hbar}{i}\frac{\partial}{\partial p_i} u_{\bq\bp}(\bx)
  \\
  \intertext{and}
  \label{eq:pu1}
  (\Hat{p}_i-p_i)u_{\bq\bp}(\bx) &=
  \Bigl(-\frac{\hbar}{i}\frac{\partial}{\partial q_i} -p_i\Bigr)u_{\bq\bp}(\bx).
\end{align}
Inserting these into \eqref{eq:step1} gives
\begin{align}
\frac{\partial \eta}{\partial t} &=
\frac{i}{\hbar}\Bigl(\bp \cdot \frac{\partial H}{\partial \bp} - H\Bigr)\eta
+ \sum_i\Bigl(\frac{\partial H}{\partial q_i} \,\frac{\partial \eta}{\partial p_i}
-\frac{\partial H}{\partial p_i} \,\frac{\partial \eta}{\partial q_i}\Bigr)
\nonumber\\
\label{eq:classwave1}
& =
\frac{i}{\hbar}\Bigl(\bp \cdot \frac{\partial H}{\partial \bp} - H\Bigr)\eta
- \{\eta, H\}
\end{align}
using the Poisson bracket \cite[Sec.~9-4]{goldstein}
\begin{equation}
  \{B,C\} = \sum_i
  \Bigl(\frac{\partial B}{\partial q_i} \, \frac{\partial C}{\partial {p_i}} -
\frac{\partial  B}{\partial {p_i}} \, \frac{\partial  C}{\partial {q_i}}\Bigr).
\end{equation}

The equation \eqref{eq:classwave1} and its derivation will be referred
to as the Schr\"odinger-Ehrenfest (\emph{SE}) equation or
approximation, because approximating the Hamiltonian as linear across
a wavepacket is often done when presenting Ehrenfest's theorem. When
we are discussing second-order terms from the expansion of $H$ we will
qualify \eqref{eq:classwave1} as the first-order SE equation.

Since $\eta(\bq,\bp)$ is the amplitude for observing the values $q$ and
$p$, the probability density on phase space is
\begin{equation}
\rho(\bq,\bp,t) = |\eta|^2  .
\end{equation}
From \eqref{eq:classwave1} the time evolution of $\rho$ is Liouville's
equation for the classical time evolution of phase space density
\cite[Sec.~9-8]{goldstein},
\begin{equation}\label{eq:Liouville}
\frac{\partial \rho}{\partial t} = -\{\rho,H\}.
\end{equation}

Note the qualitative properties of the solutions of the SE equation
\eqref{eq:classwave1}. If a classical solution $\bq(t),\bp(t)$ is
inserted, the terms with partial derivatives of $\eta$ form the total
time derivative of $\eta$ \cite[Sec.~9-5]{goldstein}. This then can be
written as
\begin{equation}
  \frac{d\eta}{dt} =
  \frac{i}{\hbar}\Bigl(\bp \cdot \frac{\partial H}{\partial \bp} - H\Bigr)\eta.
\end{equation}
Since the coefficient of $\eta$ is pure imaginary, if we followed
$\eta$ along this classical trajectory only the phase changes while
the amplitude is constant. The change in phase from time $t_0$ to
$t_1$ is
\begin{equation}
  \Delta \arg(\eta) =
  \hbar^{-1}\int_{t_0}^{t_1}\!dt\, \Bigl(\bp \cdot \frac{\partial H}{\partial \bp} - H\Bigr).
\end{equation}
The expression $\bp \cdot \partial H / \partial \bp - H$ is
numerically equal to the Lagrangian, so $\Delta \arg(\eta)$ is
$\hbar^{-1}$ times the action, specifically Hamilton's principal
function $S$ \cite[Sec.~10-1]{goldstein}, though as a function of
different arguments. Here it is a function of $\bq,\bp$ and $t$, and
has a value equal to Hamilton's principal function for the classical
trajectory that arrives at $\bq,\bp$ at time $t$ from some starting
time.

Given that the phase changes along a trajectory by the action divided
by $\hbar$, we might expect some form of action quantization from the
time-independent SE equation. We will show this in
Section~\ref{sec:bohrsommerfeld}.

\subsection{A generalization and a derivation of the KvN formulation of
  classical mechanics}
\label{sec:kvn}

The wavepackets used in the preceding expansion can be generalized to
give a derivation of the KvN formulation of classical mechanics.
Alter the wavepackets $u_{\bq\bp}$ with an additional phase $\phi(\bq,\bp,t)$,
independent of $\bx$.  The wavepackets become
\begin{equation}\label{eq:uphi}
  u_{\bq\bp}(\bx,t) = u_0(\bx-\bq)
  e^{\frac{i}{\hbar} \bp \cdot (\bx-\bq) + i\phi(\bq,\bp,t)}.
\end{equation}
The time derivative of $\eta$ changes to
\begin{equation}
\frac{\partial \eta}{\partial t} =
-i\frac{\partial \phi}{\partial t}
- \frac{i}{\hbar}\int\!dx\, [H(\bx, \Hat{\bp}, t)u_{\bq\bp}(\bx)]^* \psi(\bx,t).
\end{equation}
The expressions \eqref{eq:xu1} and \eqref{eq:pu1} change to
\begin{equation}
  \begin{aligned}\label{eq:xu1phi}
    &(x_i-q_i) u_{\bq\bp}(\bx) = \frac{\hbar}{i}\Bigl(\frac{\partial}{\partial p_i} -
    i\frac{\partial\phi}{\partial p_i}\Bigr)u_{\bq\bp}(\bx)
    \qquad\text{and} \\
    &(\hat{p}_i-p_i)u_{\bq\bp}(\bx) =
    \Bigl[-\frac{\hbar}{i}\Bigl(\frac{\partial}{\partial q_i}
      - i\frac{\partial\phi}{\partial q_i}\Bigr) - p_i\Bigr]u_{\bq\bp}(\bx).
  \end{aligned}
\end{equation}
The SE equation \eqref{eq:classwave1} becomes
\begin{equation}
  \label{eq:classwave1phi}
  \frac{\partial \eta}{\partial t} =
  \frac{i}{\hbar}\Bigl[\bp \cdot \frac{\partial H}{\partial \bp} - H
  - \hbar\Bigl(\frac{\partial \phi}{\partial t}
    + \{\phi, H\}\Bigr) \Bigr]\eta
  - \{\eta, H\}.
\end{equation}
For example, we could choose $\phi = -H(\bq,\bp)t/\hbar$, and the term in
square brackets multiplying $\eta$ would reduce to $(i/\hbar)
\bp \cdot \partial H/\partial \bp$.

To get the KvN formulation of classical mechanics, choose $\phi =
\hbar^{-1}S(\bq,\bp,t)$, with $S$ equal in value to Hamilton's principal
function as described above. Since the terms $\partial \phi / \partial
t + \{\phi, H\}$ equal the total time derivative of $\phi$ if
classical solutions are inserted for $\bq$ and $\bp$, for $\phi =
\hbar^{-1}S(\bq,\bp,t)$ these equal the Lagrangian divided by $\hbar$ and
cancel the other terms in square brackets in
Eq.~\eqref{eq:classwave1phi}, leaving
\begin{equation}
\frac{\partial \eta}{\partial t} = -\{\eta, H\}.
\label{eq:KvN}
\end{equation}
This and Liouville's equation \eqref{eq:Liouville} that follows from
it are the KvN formulation of classical mechanics
\cite{mauro2002koopman,koopman1931hamiltonian,neumann1932operatorenmethode}.
Koopman and von~Neumann postulated this rather than deriving it.

\subsection{Wavefunctions on phase space as a larger Hilbert space\label{sec:embedding}}

One might suspect that the set of wavefunctions $\eta(\bq,\bp)$ on phase
space represents a larger Hilbert space than the usual wavefunctions
$\psi(\bx)$. In the preceding this was ignored in order to give a simple
derivation of the basic equation \eqref{eq:classwave1}. In this
section the relation between the usual and larger Hilbert spaces is
described in some detail. This is mostly a review, as will be
discussed in Section~\ref{sec:prevwork}. The images $\eta(\bq,\bp)$ of
wavefunctions $\psi(\bx)$ under \eqref{eq:lift0} will be shown to form a
subspace of the set $\hC$ of $L^2$ functions on phase space. There is
a natural projection operator from $\hC$ to $\hQ$, so that an $\eta
\notin \hQ$ can be close to $\hQ$, and therefore a good approximation
to some $\psi \in \hQ$. This superspace-subspace relationship will be
constructed. While these ideas weren't strictly needed for the
preceding presentation, they will be necessary for semiclassical
quantization.

Consider a Hilbert space $\hC$ representable with complex-valued $L^2$
functions on $\mathbb{R}^{2D}$. Use $\bq,\bp$ for the $2D$ coordinates,
and $|\bq,\bp\rangle$ for joint eigenvectors of the operators for those
coordinates, scaled so that
\begin{equation}\label{eq:qpnorm}
\langle \bq',\bp'|\bq,\bp\rangle = (2\pi\hbar)^D\delta^D(\bq-\bq')\delta^D(\bp-\bp').
\end{equation}
A resolution of the identity in $\hC$ is
\begin{equation}\label{eq:Cidentity}
  \oI_\hC = \int\Bigl(\frac{dq\,dp}{2\pi\hbar}\Bigr)^D |\bq,\bp\rangle \langle \bq,\bp|.
\end{equation}
With the functions $u_{\bq\bp}(\bx)$ from \eqref{eq:udef}, construct the
set of vectors
\begin{equation}\label{eq:xconstruct}
|\bx\rangle = \int\Bigl(\frac{dq\,dp}{2\pi\hbar}\Bigr)^D u^{*}_{\bq\bp}(\bx) |\bq,\bp\rangle.
\end{equation}
Multiplying this on the left with $\langle \bq',\bp'|$, using
\eqref{eq:qpnorm}, and taking the conjugate gives
\begin{equation}\label{eq:uasxdotqp}
  u_{\bq\bp}(\bx) = \langle \bx|\bq,\bp\rangle.
\end{equation}
Multiplying \eqref{eq:xconstruct} by its conjugate and using the
self-inverting property \eqref{eq:self-invert} of the $u_{\bq\bp}$ gives
\begin{equation}
  \langle \bx|\bx'\rangle = \delta^D(\bx-\bx').
\end{equation}
Therefore these $|\bx\rangle$ can be used as the position basis for
the usual Hilbert space $\hQ$ of quantum mechanics, here as a subspace
of $\hC$. (See \cite{ban1998phase} for a procedure that starts with
$\hQ$ and constructs $\hC$ as a superspace. This might seem more
natural but is somewhat more involved.)

With these relations the expression \eqref{eq:lift0} for mapping
$\psi(\bx)$ into $\eta(\bq,\bp)$ becomes
\begin{equation}
  \label{eq:lift1}
  \eta(\bq,\bp) =
  \int \!d^D\!x\, \langle \bq,\bp|\bx\rangle \langle \bx|\psi\rangle.
\end{equation}
In these terms this is obvious since $\int \!d^Dx\, |\bx\rangle \langle
\bx|$ is a resolution of the identity in $\hQ$ and $|\psi\rangle \in
\hQ$. It then simply says $\eta(\bq,\bp) = \langle \bq,\bp|\psi\rangle$, i.e.,
$\eta(\bq,\bp)$ is $|\psi\rangle$ expressed in the $|\bq,\bp\rangle$
basis. Note that while
\begin{equation}
 \Hat{P}_\hQ \defeq \int \!d^D\!x\, |\bx\rangle \langle \bx|
\end{equation}
is a resolution of the identity in $\hQ$, within $\hC$ it is an
expression for a projection operator from $\hC$ to
$\hQ$.

Equations~\eqref{eq:self-invert} and \eqref{eq:project0} for the
mapping of $\eta(\bq,\bp)$ back to $\psi(\bx)$ become
\begin{align}
  \label{eq:self-invert1}
  \int \Bigl(\frac{dq\, dp}{2\pi\hbar}\Bigr)^D\,
  \langle \bx|\bq,\bp\rangle
  \langle \bq,\bp|\bx'\rangle &= \delta^D(\bx-\bx') \\
  \intertext{and}
  \label{eq:project1}
  \int \Bigl(\frac{dq\, dp}{2\pi\hbar}\Bigr)^D\,
  \langle \bx|\bq,\bp\rangle\langle \bq,\bp|\psi, t\rangle &= \psi(\bx,t).
\end{align}
These follow trivially from the identity operator $\oI_\hC$ expressed
in \eqref{eq:Cidentity}.

Equation~\eqref{eq:project0} expressed the fact that a wavefunction
$\psi(\bx)$ can be recovered from its mapping into $\eta(\bq,\bp)$, but we
can replace $|\psi, t\rangle \in \hQ$ in \eqref{eq:project1} with any
vector $|\eta\rangle \in \hC$ to get
\begin{equation}
  \label{eq:project2}
  \int \Bigl(\frac{dq\, dp}{2\pi\hbar}\Bigr)^D\,
  \langle \bx|\bq,\bp\rangle\langle \bq,\bp|\eta\rangle = \psi(\bx).
\end{equation}
This expresses the projection of an arbitrary vector $|\eta\rangle \in
\hC$ onto $\hQ$ in the $|\bx\rangle$ basis.

If $\hQ$ is a subspace of $\hC$, the projection operator $\Hat{P}_\hQ$
should have a null space $\hN$. The projection operator onto this null
space should be the identity minus $\Hat{P}_\hQ$, i.e., $\Hat{P}_\hN =
\oI_\hC - \Hat{P}_\hQ$, which we can write as
\begin{equation}
  \Hat{P}_\hN =
  \int \Bigl(\frac{dq\,dp}{2\pi\hbar}\Bigr)^D |\bq,\bp\rangle \langle \bq,\bp|
  - \int \!d^D\!x\, |\bx\rangle \langle \bx|.
\end{equation}
It is easy to see that $\Hat{P}_\hN |\bx\rangle = 0 $ using
\eqref{eq:xconstruct}, the construction of $|\bx\rangle$.  Similarly we
can show that
\begin{multline}
  \langle \bq,\bp|\Hat{P}_\hN |\bq_0,\bp_0 \rangle
  = (2\pi\hbar)^D \, \delta^D(\bq-\bq_0)\delta^D(\bp-\bp_0) \\
  - \int \!d^D\!x\,u^*_{\bq\bp}(\bx)u_{\bq_0\bp_0}(\bx).
\end{multline}
We can work this out for a particular choice of the $u_{\bq\bp}(\bx)$, such
as Gaussians, and verify that it is non-zero. Thus none of the $|\bx
\rangle$ have components in the null space, while all $|\bq_0,\bp_0
\rangle$ have non-zero projection onto the null space, so none lie in
$\hQ$.

It is perhaps obvious that $\bq$ and $\bp$ are not exact position and
momentum eigenvalues, but particular kinds of approximations. There
are, however, relationships between the position and momentum
operators on $\hC$ and $\hQ$. On $\hC$ the natural choices for
classical mechanics are
\begin{align}
  \label{eq:positionP}
  \Hat{q}_{\hC,i} &\defeq \int \Bigl(\frac{dq\,dp}{2\pi\hbar}\Bigr)^D
  \, |\bq\bp \rangle q_i \langle \bq\bp|
  \qquad\text{and} \\
  \label{eq:momentumP}
  \Hat{p}_{\hC,i} &\defeq \int \Bigl(\frac{dq\,dp}{2\pi\hbar}\Bigr)^D
  \, |\bq\bp \rangle p_i \langle \bq\bp|.
\end{align}
Clearly these commute, so their action on elements of $\hQ$ can't be
the same as the usual quantum position and momentum
operators. However, we can show that the projections onto $\hQ$,
namely $\Hat{P}_\hQ \Hat{q}_{\hC,i} \Hat{P}_\hQ$ and $\Hat{P}_\hQ
\Hat{p}_{\hC,i} \Hat{P}_\hQ$, are the usual quantum operators
$\Hat{x}_{\hQ,i}$ and $\Hat{p}_{\hQ,i}$. For example, from the
definition of \(\Hat{P}_\hQ\) above
\begin{multline}\label{eq:pc-project}
  \Hat{P}_\hQ \Hat{p}_{\hC,i} \Hat{P}_\hQ = \\
  \int\!d^D\!x\,d^D\!\bx'\, |\bx\rangle
  \Bigl[ \int \Bigl(\frac{dq\,dp}{2\pi\hbar}\Bigr)^D \langle \bx|\bq\bp \rangle
  p_i
  \langle \bq\bp|\bx'\rangle \Bigr]
  \langle \bx'|.
\end{multline}
The factor in square brackets is
\begin{align}
  \int \Bigl(\frac{dq\,dp}{2\pi\hbar}\Bigr)^D \,u_0(\bx-\bq) u^{*}_0(\bx'-\bq)
  \,p_i\, e^{\frac{i}{\hbar}\bp_i (\bx_i - \bx'_i)} \nonumber \\
  = \int\!d^D\!q \,u_0(\bx-\bq) u^{*}_0(\bx'-\bq)
  \,\frac{\hbar}{i} \frac{\partial}{\partial x_i}\delta^D(\bx-\bx') \nonumber \\
  = \frac{\hbar}{i} \frac{\partial}{\partial x_i}\delta^D(\bx-\bx').
\end{align}
Re-inserting into \eqref{eq:pc-project} gives
\begin{equation}
  \Hat{P}_\hQ \Hat{p}_{\hC,i} \Hat{P}_\hQ = \int\!d^D\!x\,|\bx\rangle
  \frac{\hbar}{i}\frac{\partial}{\partial x_i} \langle \bx| = \Hat{p}_{Q,i}.
\end{equation}

More general classical observables are functions of $\bq,\bp$, so the
corresponding operators in $\hC$ are 
\begin{equation}
  \label{eq:operatorC}
  \Hat{O}_{\hC} =
  \int \Bigl(\frac{dq\,dp}{2\pi\hbar}\Bigr)^D \, |\bq\bp \rangle O(\bq,\bp) \langle \bq\bp|.
\end{equation}
In general the projection of this onto $\hQ$ is not the same as
replacing $\bq,\bp$ in $O(\bq,\bp)$ with $\Hat{\bx}, \Hat{\bp}$ to get $O(\Hat{\bx},
\Hat{\bp})$.

\subsection{Relation to previous work\label{sec:prevwork}}

Much of Section~\ref{sec:embedding} and parts of
Section~\ref{sec:cfromq} are in the previous literature on phase space
wavefunctions. In this section the similarities and differences are
presented in more detail.

In the earliest of these papers \cite{torres1990quantum}, Torres-Vega
and Frederick use wavepackets of the same form as \eqref{eq:udef},
relate them to the inner product of phase space basis vectors and
coordinate basis vectors as in \eqref{eq:uasxdotqp}, map a
wavefunction in the coordinate representation into the phase space
representation with \eqref{eq:lift0}, and use the expressions
\eqref{eq:xu1} for $u_{\bq\bp}$ multiplied by $x_i-q_i$ and
$\Hat{p}_i-q_i$, though all in their notation. They then proceed to
express Schr\"odinger's equation in phase space for Hamiltonians of
the form $\|\bp\|^2/2m + V(\bq)$, solve for energy eigenstates of the
harmonic oscillator, and examine the time evolution of coherent
states. They note the resemblance of the resulting equation for the
time evolution of the phase space density of coherent states to
Liouville's equation.

Harriman \cite{harriman1994quantum} has a similar presentation with a
more general approach to the matrix elements or kernels
\eqref{eq:uasxdotqp}. Early in the paper he narrows his discussion to
Gaussian kernels or coherent states of the harmonic oscillator, for
the sake of treating position and momentum similarly. This choice
results in an eigenvalue equation satisfied by any element of $\hQ$,
which discriminates between elements of $\hQ$ and other elements of
$\hC$. He then proceeds to discuss the phase space forms of various
position-space wavefunctions and momentum-space wavefunctions.

The SE equation \eqref{eq:classwave1} was not derived in that previous
work. This approach of deriving the classical limit directly, rather
than first formulating quantum mechanics on phase space, allows
considerable simplification, more general Hamiltonians, and more
general wavepacket families. To the author's knowledge this and the
results in the remainder of the paper are new.

\subsection{Second-order terms}\label{sec:essecondorder}

In this section the summation convention is used. The second-order
terms dropped from \eqref{eq:step1} are
\begin{equation}
  \begin{split}
    -\frac{i}{\hbar} \biggl(
 &\frac{1}{2}\frac{\partial^2 H}{\partial p_i \partial p_j}
    \int\!dx\bigl[(\Hat{p}_i-p_i) (\Hat{p}_j-p_j) u_{\bq,\bp}(\bx)\bigr]^{*} \psi \\
+ & \frac{\partial^2 H}{\partial q_i \partial p_j}
    \int\!dx\Bigl\{ \frac{1}{2} [x_i-q_i, \Hat{p_j}-p_j]_{+} u_{\bq,\bp}(\bx)\Bigr\}^{*} \psi \\
+ & \frac{1}{2}\frac{\partial^2 H}{\partial q_i \partial q_j}
    \int\!dx\bigl[(x_i-q_i) (x_j-q_j) u_{\bq,\bp}(\bx)\bigr]^{*} \psi \biggr).
\end{split}
\label{eq:step1_order2}
\end{equation}
(Here the anticommutator $\frac{1}{2}[x_i-q_i, \Hat{p}_j-p_j]_{+}$ is used to make
the operator Hermitian.) Like \eqref{eq:xu1} and \eqref{eq:pu1}, it
can be shown that
\begin{multline}
  \label{eq:pu2}
  (\Hat{p_i}-p_i) (\Hat{p_j}-p_j) u_{\bq,\bp}(\bx) =
  \\
  \Bigl(-\frac{\hbar}{i}\frac{\partial}{\partial q_i}-p_i\Bigr)
  \Bigl(-\frac{\hbar}{i}\frac{\partial}{\partial q_j}-p_j\Bigr) u_{\bq,\bp}(\bx),
\end{multline}
\begin{multline}
\label{eq:pxu2}
\frac{1}{2}[x_i-q_i, \Hat{p}_j-p_j]_{+} u_{\bq,\bp}(\bx) = \\
\Bigl[\bigl(-\frac{\hbar}{i}\frac{\partial}{\partial q_j}-p_j\bigr)\frac{\hbar}{i}\frac{\partial}{\partial p_i}
  - \frac{\hbar}{2i}\delta_{i,j}\Bigr] u_{\bq,\bp}(\bx),
\end{multline}
and
\begin{multline}
  \label{eq:xu2}
  (x_i-q_i) (x_j-q_j) u_{\bq,\bp}(\bx) =
  -\hbar^2 \frac{\partial^2}{\partial p_i \partial p_j} u_{\bq,\bp}(\bx).
\end{multline}
Inserting these into \eqref{eq:step1_order2} and then adding to
\eqref{eq:classwave1} gives
\begin{multline}
  \frac{\partial \eta}{\partial t} =
  \frac{i}{\hbar}\Bigl(p_i\frac{\partial H}{\partial p_i} -H\Bigr)\eta - \{\eta,H\}
  \\
  + \Bigl[
    \frac{-i}{2\hbar}p_i p_j \frac{\partial^2 H}{\partial p_i \partial p_j}
    - \frac{1}{2}\frac{\partial^2 H}{\partial q_i \partial p_i}\Bigr] \eta
  \\
  + p_i\Bigl[\frac{\partial^2 H}{\partial p_i \partial p_j}
    \frac{\partial \eta}{\partial q_j}
    - \frac{\partial^2 H}{\partial p_i \partial q_j}
    \frac{\partial \eta}{\partial p_j} \Bigr] \\
  +\frac{i\hbar}{2}\Bigl[
    \frac{\partial^2 H}{\partial p_i \partial p_j}
    \frac{\partial^2 \eta}{\partial q_i \partial q_j}
    - 2 \frac{\partial^2 H}{\partial p_i \partial q_j}
    \\
    \frac{\partial^2 \eta}{\partial q_i \partial p_j}
    + \frac{\partial^2 H}{\partial q_i \partial q_j}
    \frac{\partial^2 \eta}{\partial p_i \partial p_j} \Bigr].
  \label{eq:classwave2}
\end{multline}
The new terms couple the phase and amplitude, and thus the time
derivative of the density $\rho = \lvert \eta \rvert^2$ no longer
involves only terms with $\rho$ and its derivatives.

\section{Semi-classical quantization\label{sec:bohrsommerfeld}}

Here we first derive the time-independent SE equation. We then use
this to derive the Bohr-Sommerfeld quantization formula $J_i=h n_i$
\cite[Section~I.15]{messiah1970quantum} for separable systems. This is
followed by a discussion of the exact harmonic oscillator energy
eigenfunctions mapped into phase space. These obey the first-order SE
equation with the Bohr-Sommerfeld result $E=\hbar \omega n$. We then
show that the ground-state term $\hbar \omega / 2$ comes from the
second-order terms in the SE approximation.

\subsection{The approximate time-independent Schr\"odinger-Ehrenfest equation}

We can derive equations for semi-classical quantization using the
time-independent form of the SE equation
Eq.~\eqref{eq:classwave1}. Assume that the Hamiltonian doesn't depend
explicitly on time, and consider an energy eigenstate
$\psi(\bx)\exp(-iEt/\hbar)$ in $\hQ$. The partial time derivative of
$\eta$ is $-\frac{i}{\hbar}E\eta$. Inserting this into the SE equation
\eqref{eq:classwave1} gives
\begin{equation}\label{eq:classwave1timeind}
  E\eta = 
  \Bigl(H - p \cdot \frac{\partial H}{\partial p}\Bigr)\eta
  + \frac{\hbar}{i}\{\eta, H\}.
\end{equation}
Represent $\eta$ with an amplitude and phase,
\begin{equation}
\eta(\bq,\bp,t) = A(\bq,\bp) \exp\Bigl\{\frac{i}{\hbar}\bigl[\beta(\bq,\bp) - Et\bigr]\Bigr\},
\end{equation}
where $A$ and $\beta$ are real-valued.  The real and
imaginary parts of \eqref{eq:classwave1timeind} become, after dividing
out some factors,
\begin{align}
  \label{eq:amplitude}
  0 &= \{A, H\}
  \qquad\qquad\text{and} \\
  \label{eq:phase}
  E &=
  H - p \cdot \frac{\partial H}{\partial p}
  + \{\beta, H\}.
\end{align}
The amplitude and phase are independent, as pointed out
in \cite{mauro2002koopman} for the KvN formulation.

If a phase-space function $B$ has no explicit time dependence and
classical solutions $q(t), p(t)$ are inserted, $\{B, H\}$ equals
$dB/dt$, so \eqref{eq:amplitude} and \eqref{eq:phase} relate the
values of $A$ or $\beta$ along any classical trajectory, including any
manifold covered by a trajectory. In particular \eqref{eq:amplitude}
requires $A$ to be constant on such a manifold. Since they follow
classical trajectories the expressions \eqref{eq:amplitude} and
\eqref{eq:phase} can't relate $A$ or $\beta$ across different values
of the constants of the motion, in particular $H$.

One might try to simplify \eqref{eq:phase} by using the KvN equation
\eqref{eq:KvN}, derived by adding the phase $S(\bq,\bp,t)/\hbar$ to
the $u_{\bq\bp}$. However, introducing an extra phase $\phi$ changes
the partial time derivative of $\eta$ to $(-i/\hbar)(E + \hbar
\partial \phi /\partial t)\eta$. Inserting this into
\eqref{eq:classwave1phi}, the terms involving $\partial \phi /\partial
t$ cancel, leaving $\{\phi,H\}\eta$ as the only term involving
$\phi$. Therefore using $\phi = S(\bq,\bp,t)/\hbar$ does not give the
total time derivative of the action, and does not cancel the terms $H
- p \cdot\partial H / \partial p$.

There is nothing in the preceding that requires $H = E$.  In
Section~\ref{sec:embedding} it was shown that wavefunctions on phase
space represent a Hilbert space $\hC$, and that the usual Hilbert
space $\hQ$ is a subspace of $\hC$. In Appendix~\ref{app:HapproxE} it
is shown that solutions with $H \neq E$ approach the null space of the
projection from $\hC$ onto $\hQ$ as $\lvert E - H \rvert$ grows, and
therefore they cannot be good approximations to an eigenfunction in
$\hQ$. The scale for which $\lvert E - H \rvert$ is small enough to
contribute is related to the system, the spatial width $\sigma$ of the
wavepackets $u_{\bq\bp}$, and $\hbar$. Points with $H=E$ make the
dominant contribution. Assuming the wavefunction has support only for
$H(\bq,\bp)=E$, there are three equations for an approximate energy
eigenstate and eigenvalue $E$: The equation \eqref{eq:amplitude} for
the amplitude, $H=E$, and
\begin{equation}
  \label{eq:phase2}
  0 =
  - p \cdot \frac{\partial  H}{\partial p} + \{\beta, H\}.
\end{equation}

\subsection{Bohr-Sommerfeld quantization}

Quantization follows from the time-independent SE equation
\eqref{eq:phase2} and the continuity of $\eta$, which in turn requires
that the phase $\beta/\hbar$ changes by an integer multiple of $2\pi$
as a closed path is followed on a torus given by particular values of
the constants of motion.  For a separable system we will compute the
change of phase $\beta/\hbar$ around suitable closed paths. These must
be integer multiples of $2\pi$ as described above, and the result is
Bohr-Sommerfeld quantization.

First, the preceding was all in Cartesian coordinates, but in general
a system is separable in some other coordinates. So we need to see how
the equation \eqref{eq:phase2} governing $\beta$ changes. It is shown
in Appendix~\ref{app:pdqinv} that $p \cdot \partial H / \partial p$ is
invariant under coordinate transformations, so Eq.~\eqref{eq:phase2}
is also invariant, since the Poisson bracket is a canonical invariant.

In separable coordinates $\bq,\bp$, compute the indefinite integral of
\eqref{eq:phase2} to get an expression for $\beta$. Consider a
classical trajectory parameterized by a time-like variable $t$. Using
$\partial H / \partial p_i = \Dot{q}_i$ and $\{\beta, H\} =
d\beta(\bq,\bp)/dt$, integrate \eqref{eq:phase2} over $t$ to get
\begin{equation}
\int \! dt\, p \cdot \Dot{q} = \beta.
\end{equation}
The left-hand side is Hamilton's characteristic function, $W$
\cite[Sec.~10-3]{goldstein}, so $\beta$ is numerically equal to
$W$. They are, however, functions of different variables. The
arguments of $W$ are the coordinates $q$ and the $D$ constants of the
motion $\alpha$, but not the momenta, while the arguments of $\beta$
are the coordinates and momenta but not $\alpha$. $W$ and $\beta$ can
be equal since $\alpha$ and $\bq,\bp$ are not independent. We can express
the constants of motion as functions $c(\bq,\bp)$, with values $\alpha =
c(\bq,\bp)$ when $\bq,\bp$ satisfies those constants of motion. With $c$
inserted for the $\alpha$ arguments we can write
\begin{equation}\label{eq:Qbeta}
  \beta(q, p) = W\bigl(q, c(q, p)\bigr).
\end{equation}
There is another important difference: $W$ is multiple-valued, since
$\partial W(q,\alpha) / \partial q_i = p_i$, and a particular value of
$q$ does not completely fix the momenta $p$. For example, with a
harmonic oscillator in one dimension a given value of $q$ fixes the
magnitude of $p$ but not its sign. In general there must be different
$W$'s for different parts of the trajectory. In $\beta$ these are
disambiguated since $\beta$ depends on both $q$ and $p$. Thus at any
particular point in phase space \eqref{eq:Qbeta} holds with one of the
values of $W$, while $\beta$ is a single-valued function on all of
phase space.

We can use \eqref{eq:Qbeta} to evaluate the change in $\beta$ as the
system follows a closed path that is the intersection of the torus for
given values of the constants of the motion and the plane of a
particular $q_i,p_i$ pair with the other variables held fixed. We'll
show that this intersection is followed by a path satisfying
$dq_i/dt_i = \partial H / \partial p_i$ and $dp_i/dt_i = -\partial H /
\partial q_i$ with parameter $t_i$, and all $q_j,p_j$ for $j\neq i$
held fixed, i.e., it preserves the constants of motion, even though it
is not a solution of the full set of equations of motion. Then
\eqref{eq:Qbeta} will determine how $\beta$ changes along this path.

For brevity use
\begin{equation}
  \{B, C\}_i = \frac{\partial B}{\partial q_i} \frac{\partial
    C}{\partial p_i} - \frac{\partial B}{\partial p_i}
  \frac{\partial C}{\partial q_i},
\end{equation}
for the terms in the Poisson bracket, so $\{B,C\}=\sum_i\{B,C\}_i$.
In Appendix~\ref{app:ciHj} it is shown that, for all $i$ and $j$,
\begin{equation}\label{eq:ciHj}
  \{c_i, H\}_j = 0.
\end{equation}
On the path mentioned above, with a specific $q_i,p_i$ pair completing
an orbit while $q_j,p_j$ for all $j \neq i$ are fixed, because of
\eqref{eq:ciHj} all of the constants of the motion are satisfied, so
the path lies in the solution torus. The derivative of $\beta$ with
respect to $t_i$ is
\begin{align}
  \frac{d\beta}{dt_i} &= \frac{\partial \beta}{\partial q_i} \frac{dq_i}{dt_i}
  + \frac{\partial \beta}{\partial p_i} \frac{dp_i}{dt_i} \nonumber \\
  \label{eq:dbetadt1}
  &= \frac{\partial \beta}{\partial q_i} \frac{\partial H}{\partial p_i}
  - \frac{\partial \beta}{\partial p_i} \, \frac{\partial H}{\partial q_i}.
\end{align}
Use the relation \eqref{eq:Qbeta} between $\beta$ and $W$ to get
\begin{align}
   \nonumber
   \frac{\partial \beta}{\partial q_i} &=
   \frac{\partial W}{\partial q_i}
   + \sum_j \frac{\partial W}{\partial \alpha_j}\frac{\partial c_j}{\partial q_i} \\
   \label{eq:dbdq_W}
  &= p_i  + \sum_j \frac{\partial W}{\partial {\alpha_j}}\frac{\partial c_j}{\partial q_i}
  \qquad\text{and}\\
   \label{eq:dbdp_W}
   \frac{\partial \beta}{\partial p_i} &=
   \sum_j \frac{\partial W}{\partial \alpha_j}\frac{\partial c_j}{\partial p_i}.
\end{align}
Inserting these into \eqref{eq:dbetadt1} and using \eqref{eq:ciHj}
gives
\begin{align}
  \frac{d\beta}{dt_i}
  \nonumber
  &= p_i \frac{\partial H}{\partial p_i}
  + \sum_j \frac{\partial W}{\partial \alpha_j} \{c_j,H\}_i
  \\
  \label{eq:dbdt_pdHdp}
  &= p_i \frac{\partial H}{\partial p_i}.
\end{align}
Define $\Delta_i\beta$ as the change in $\beta$ after completing one
orbit on this path. Integrating \eqref{eq:dbdt_pdHdp} for one
period of $t_i$ gives
\begin{equation}
  \Delta_i\beta = \oint p_i\,dq_i = J_i.
\end{equation}
Since the phase $\beta/\hbar$ can only change by an integer multiple
of $2\pi$, this requires
\begin{equation}
  J_i = n_ih,
\end{equation}
i.e., Bohr-Sommerfeld quantization.

\subsection{The harmonic oscillator}\label{sec:hoexample}

As a simple example, consider the one-dimensional harmonic oscillator.
We'll show that the exact energy eigenstates mapped into $\hC$ using a
particular set of $u_{\bq\bp}$ obey the first-order SE equation if the
energy eigenvalue is the Bohr-Sommerfeld value $E_n = \hbar \omega
n$. The ground-state energy term $\hbar \omega / 2$ is given by the
second-order terms in the SE approximation.

These eigenstates are
\begin{equation}
  \psi_n(x) = \Bigl(\frac{m\omega}{2^{2n}(n!)^2\pi\hbar}\Bigr)^{1/4}
    H_n\Bigl(\sqrt{\frac{m\omega}{\hbar}} x\Bigr) e^{-m\omega x^2 / 2\hbar}
\end{equation}
where $H_n$ is the $n$-th Hermite polynomial with normalization such
that the highest-order term has coefficient $2^n$.  We'll map this
into $\eta(\bq,\bp)$ using the wavepackets
\begin{equation}
  u_{\bq\bp}(x) = (2\pi\sigma^2)^{-1/4}
  \exp\Bigl[ -\frac{(x-q)^2}{4\sigma^2} + \frac{i}{\hbar}p(x-q)\Bigr]
\end{equation}
The mapping into $\hC$ is
\begin{multline}
  \eta_n(\bq,\bp) = \Bigl(\frac{m\omega}{2^{2n+1}(n!)^2\pi^2\hbar\sigma^2}\Bigr)^{1/4}
  \int\! dx\, H_n\Bigl(\sqrt{\frac{m\omega}{\hbar}} x \Bigr) \\
  \times \exp\Bigl(
    -\frac{m\omega x^2}{2\hbar}
    -\frac{(x-q)^2}{4\sigma^2} - \frac{i}{\hbar}p(x-q)
  \Bigr).
\end{multline}
Defining
\begin{equation}\label{eq:zetadef}
  \zeta \defeq \sqrt{\frac{\hbar}{m\omega}}
\end{equation}
shortens this a little, to
\begin{multline}\label{eq:eta_x}
  \eta_n(q, p) = \frac{1}{\sqrt{2^{(2n+1)/2} n! \pi\zeta\sigma}}
  \int\! dx\, H_n(x/\zeta)
  \\
  \times
  \exp\Bigl(
    -\frac{x^2}{2\zeta^2}
    -\frac{(x-q)^2}{4\sigma^2} - \frac{i}{\hbar}p(x-q)
  \Bigr).
\end{multline}
Define further
\begin{equation}\label{eq:alphadef}
  \alpha \defeq \sqrt{\frac{4\sigma^2}{2\sigma^2 + \zeta^2}}.
\end{equation}
Letting $y \defeq x/\alpha\zeta$ the wavefunction \eqref{eq:eta_x} becomes
\begin{multline}
  \eta_n(q, p) =
  \sqrt{\frac{\alpha^2\zeta}{2^{(2n+1)/2}n!\pi\sigma}}
  \\
  \times\exp\Bigl(
    - \frac{\alpha^2}{8\sigma^2} q^2
    + \frac{i\alpha^2 p q}{2\hbar}
    -   \frac{\alpha^2\zeta^2 p^2}{4\hbar^2}
  \Bigr) \\
  \times
  \int\! dy\, H_n(\alpha y)
  \exp\Bigl(
  -\Bigl[y - \frac{\alpha\zeta}{2}\Bigl(\frac{q}{2\sigma^2} - \frac{i}{\hbar}p\Bigr)\Bigr]^2
  \Bigr).
\end{multline}
We can use the result \cite[entry~7.374.8]{gandr}
\begin{multline}\label{eq:gandrres}
  \int \! dx \, e^{-(x-z)^2} H_n(\alpha x) =
  \\
  \pi^{1/2} (1 - \alpha^2)^{n/2}
  H_n\Bigl(\frac{\alpha z}{\sqrt{1 - \alpha^2}}\Bigr)
\end{multline}
to get
\begin{multline}\label{eq:eta0}
  \eta_n(\bq,\bp) =
  \sqrt{\frac{\alpha^2\zeta}{2^{(2n+1)/2}n!\sigma}}
  (1 - \alpha^2)^{n/2}
  \\
  \times
  H_n\Bigl[\frac{\alpha^2 \zeta}{2\sqrt{1 - \alpha^2}}\Bigl(\frac{q}{2\sigma^2} - \frac{i}{\hbar}p\Bigr)\Bigr]
  \\
  \times
  \exp\Bigl[
  - \frac{\alpha^2}{4}
  \Bigl(\frac{q^2}{2\sigma^2}
  - 2 \frac{i}{\hbar} p q
  + \frac{\zeta^2 p^2}{\hbar^2}\Bigr)
  \Bigr].
\end{multline}
For $\alpha > 1$ we can replace $\sqrt{1 - \alpha^2}$ with
$i\sqrt{\alpha^2 - 1}$.

The rest is greatly simplified if we let $\alpha$ approach $1$, or
equivalently to let $2\sigma^2$ approach $\zeta^2$.  The resulting
wavepackets $u_{\bq\bp}$ are the coherent states for this Hamiltonian.
The factor $(1 - \alpha^2)^{-1/2}$ in the argument to $H_n$ goes to
infinity, so $H_n$ will go to infinity as $(1 - \alpha^2)^{-n/2}$, but
$H_n$ is multiplied by $(1 - \alpha^2)^{n/2}$. The leading term then
is constant in $1 - \alpha^2$ while all sub-leading terms go to zero,
so
\begin{multline}
  (1 - \alpha^2)^{n/2}
  H_n\Bigl[\frac{\alpha^2 \zeta}{2\sqrt{1 - \alpha^2}}\Bigl(\frac{q}{2\sigma^2} - \frac{ip}{\hbar}\Bigr)\Bigr]
  \\
  \rightarrow
  (1 - \alpha^2)^{n/2} 2^n \Bigl[\frac{\alpha^2 \zeta}{2\sqrt{1 - \alpha^2}}\Bigl(\frac{q}{2\sigma^2} - \frac{ip}{\hbar}\Bigr)\Bigr]^n
\end{multline}
\begin{equation}
  = \Bigl(\frac{q}{\zeta} - \frac{ip \zeta}{\hbar} \Bigr)^n
\end{equation}
With this \eqref{eq:eta0} becomes
\begin{multline}\label{eq:eta_n_zeta}
 \eta_n(q, p) =
  \frac{1}{\sqrt{2^{n}n!}}
  \Bigl(\frac{q}{\zeta} - \frac{i}{\hbar}\zeta p\Bigr)^n
  \\
  \times
  \exp\Bigl[
  - \frac{1}{4}
  \Bigl(\frac{q^2}{\zeta^2}
  - 2 \frac{i}{\hbar} p q
  + \frac{\zeta^2 p^2}{\hbar^2}\Bigr)
  \Bigr]
\end{multline}
Inserting the definition of $\zeta$ gives
\begin{multline}\label{eq:eta_n}
  \eta_n(\bq,\bp) =
  \frac{1}{\sqrt{n!}} \Bigl(\frac{m\omega}{2\hbar}\Bigr)^{n/2}
  \Bigl(q - \frac{ip}{m\omega}\Bigr)^n
  \\
  \times
  \exp\Bigl[
  - \frac{1}{2\hbar\omega}
  \Bigl(\frac{m\omega^2 q^2}{2}
  - i \omega p q
  + \frac{p^2}{2m}\Bigr)
  \Bigr].
\end{multline}

Let us insert this into the first-order SE equation,
\eqref{eq:classwave1timeind}. The Poisson bracket is
\begin{equation}
  \{\eta, H\} =
  \frac{i}{\hbar} \Bigl (
    \hbar \omega n
    + \frac{p^2}{2 m}
    - \frac{m \omega^2 q^2}{2}
\Bigr) \eta.
\end{equation}
The first-order time-independent SE equation is
\begin{align}
  E\eta &= \Bigl( H - p\frac{\partial H}{\partial p}\Bigr) \eta
  -i\hbar\{\eta, H\}
  \\
  &= 
  \hbar \omega n \eta.
\end{align}
This lacks the ground-state energy term, $\hbar\omega/2$, which must
come from higher-order terms in this approximation. For the harmonic
oscillator the only higher-order terms are second-order.

For a one-dimensional system with $\partial^2H /\partial q \partial p
= 0$ the second order form for the
time-independent SE equation is, from Eq.~\eqref{eq:classwave2},
\begin{multline}
  E \eta =
  \Bigl(H - p\frac{\partial H}{\partial p} \Bigr) \eta - i\hbar \{\eta,H\}
  \\
  + \frac{1}{2}p^2 \frac{\partial^2 H}{\partial p^2}
  + i\hbar p\frac{\partial^2 H}{\partial p^2}
    \frac{\partial \eta}{\partial q}
  \\
  -\frac{\hbar^2}{2}\Bigl[
    \frac{\partial^2 H}{\partial p^2}
    \frac{\partial^2 \eta}{\partial q^2}
    + \frac{\partial^2 H}{\partial q^2}
    \frac{\partial^2 \eta}{\partial p^2} \Bigr].
  \label{eq:classWaveTInd1DSimple}
\end{multline}

For the harmonic oscillator these terms are
\begin{multline}
  E \eta =
  -\Bigl(\frac{p^2}{2m} - \frac{m\omega^2 q^2}{2}\Bigr)\eta
  + \frac{\hbar}{i} \Bigl(\frac{p}{m}\frac{\partial \eta}{\partial q}
  - m\omega^2 q\frac{\partial \eta}{\partial p}\Bigr) \\
  + \frac{p^2}{2m} \eta
  + i\hbar \frac{p}{m} \frac{\partial \eta}{\partial q}
  - \frac{\hbar^2}{2m}
  \frac{\partial^2 \eta}{\partial q^2}
  - \frac{m\hbar^2 \omega^2}{2}
  \frac{\partial^2 \eta}{\partial p^2}.
  \label{eq:ho1_classwave2tind_HO}
\end{multline}

With the exact eigenfunction \eqref{eq:eta_n} the second-order terms
(the second line of \eqref{eq:ho1_classwave2tind_HO}) equal
\(
  (\hbar \omega /2)  \eta.
\)
So at least in this case the terms to first order give the
Bohr-Sommerfeld result, without a ground-state energy term, while the
latter comes from the second-order terms.

\section{Discussion\label{sec:discussion}}

The first-order Schr\"odinger-Ehrenfest (SE) equation
\eqref{eq:classwave1} gives classical mechanics in a fairly intuitive
form: Wavefunctions on phase space evolve in time by being transported
along classical trajectories, with the amplitude remaining constant
while the phase changes by the action divided by $\hbar$. This applies
to any wavefunction, not just wavepackets. The squared-magnitude of
the corresponding wavefunction in phase space is a plausible
probability density, and obeys Liouville's equation, in this
approximation.

This is a derivation of the KvN formulation of classical mechanics
(Section~\ref{sec:kvn}), which previously was postulated but not
derived. There is a difference, since the phase plays essentially no
role in the KvN formulation, and is unchanging along a
trajectory. Here its form is necessary to project correctly back to
the usual Hilbert space $\hQ$, and its change along a trajectory is
important for quantization.

The version of semiclassical quantization derived here has an
intuitive interpretation as standing waves in phase space on the torus
covered by a classical trajectory. For separable systems the
Bohr-Sommerfeld formula follows, including its lack of a zero-point
term. For the case of the harmonic oscillator and its exact energy
eigenstates it was shown that the ground-state energy term comes from
the second-order terms of the SE approximation. We have not yet given
a procedure for computing zero-point terms from the second-order
approximation.

We can compare the time-independent SE approximation's picture of
quantization with Einstein-Brillouin-Keller (EBK) quantization
\cite{keller58, stone05}. There, the quantization formula arises from
the wavefunction ansatz $\psi =
\sum_kA_k\exp(\frac{i}{\hbar}S_k)$. The use of multiple terms differs
from the form $\psi = A \exp(\frac{i}{\hbar}S)$ given in the
introduction.  In \cite[Section~VI.4]{messiah1970quantum} an odd
feature of the single-term version is noted, namely that $\nabla S$
gives only one value for the momentum at each position in
configuration space. The multiple terms in the EBK ansatz are used
precisely to represent all of the possible values of momentum at each
position. These terms are conceived of as wavefunctions on different
parts of a torus. It is the matching of the phases of these terms at
the boundaries of the classically-accessible region, or edges of the
torus, that gives zero-point terms.  This matching of terms can also
be attributed to caustics, which are an artifact of representing the
state in configuration space
\cite{littlejohn1986semiclassical}. Representing the wavefunction in
momentum space, for example, gives a different approximation to the
wavefunction, since matching must now be done at the momentum turning
points instead of the position turning points. The matching of terms
can only be done for separable systems, otherwise there is a
continuous set of momenta for a given position, and there is no method
to relate the terms in this continuous set.

In the SE approximation the torus appears naturally as the phase-space
torus covered by a classical trajectory. Because the treatment is in
phase space the wavefunction does not have multiple terms, so no
matching is required. In particular it still applies to non-separable
systems.

Applying the SE approximation to non-separable systems remains to be
done. As usual solving those equations is difficult. At best it seems
likely that some of the eigenvalues can be obtained by integrating the
phase part \eqref{eq:phase2} of the SE equation on any periodic orbits
that might be known.

\appendix

\section{\texorpdfstring{$H\approx E$}{H\approx E} for eigenstates close to
  \texorpdfstring{$\hQ$}{Q}}
\label{app:HapproxE}

Here it is shown that the solutions of \eqref{eq:classwave1timeind}
approach the null space of the projection onto $\hQ$ as $H$ is farther
from $E$. The projection will be expressed in the $|\bq,\bp\rangle$
basis. Since $\hbar$ is not zero, solutions with $H \neq E$ can
contribute to an energy eigenfunction in $\hQ$, but only if $H$ is
sufficiently close to $E$.

Starting with an eigenfunction $\eta$, compute the projection onto
$\hQ$ by applying \eqref{eq:project0}, the mapping from $\eta$ to
$\psi$, and then \eqref{eq:lift0}, the mapping back to the $\lvert
\bq,\bp\rangle$ basis. Let the dimensionality of the phase space be
$2D$. The projection is
\begin{align}
  \Tilde{\eta}(\bq,\bp) &\defeq \!
  \int \! \Bigl(dx\frac{dq'\,dp'}{2\pi\hbar}\Bigr)^D
  u^{*}_{\bq\bp}(\bx)u_{\bq'\bp'}(\bx)
  \eta(\bq',\bp') \nonumber
  \\
  \label{eq:kernelprojeta}
  &= \int\Bigl(\frac{dq'\,dp'}{2\pi\hbar}\Bigr)^D K(\bq,\bp,\bq',\bp')
  \eta(\bq',\bp'),
\end{align}
where the kernel $K$ is
\begin{multline}\label{eq:Kdef}
  K(\bq,\bp,\bq',\bp') = \int\!d^Dx\, u^{*}_{\bq\bp}(\bx)
  u_{\bq'\bp'}(\bx) \\
  = \int\!d^Dx\, u^{*}_0(\bx-\bq) u_0(\bx-\bq') \\
  \times \exp\bigl(i\hbar^{-1}[-\bp\cdot(\bx-\bq) + \bp'\cdot(\bx-\bq')] \bigr).
\end{multline}
Remember that $\bq$ is a vector of Cartesian coordinates. The exact
form of $K$ depends on the form of $u_0$. For convenience use the
Gaussian
\begin{equation}
  u_0(\bx) \approx
  \frac{1}{(2\pi\sigma^2)^{1/4}} e^{-\|\bx\|^2/4\sigma^2}.
\end{equation}
If $u_0$ is not a Gaussian, this can be viewed as an approximation,
where we write $u_0(\bx)$ as $\exp(\log u_0(\bx))$ and replace $\log
u_0(\bx)$ with its second-order Taylor expansion around the $\bx$ that
maximizes it. This assumes that $u_0$ is sufficiently
well-behaved. With this form the kernel works out to
\begin{multline}\label{eq:approxkernel}
  K(\bq,\bp,\bq',\bp') = \exp\Bigl\{ -\frac{\|\bq'-\bq\|^2}{8\sigma^2}
  - \frac{\|\bp'-\bp\|^2}{2\hbar^2\sigma^{-2}} \\
  - \frac{i(\bp'+\bp)\cdot(\bq'-\bq)}{2\hbar} \Bigr\}.
\end{multline}

Evaluate the integral \eqref{eq:kernelprojeta} for $\bq,\bp$ on the
manifold satisfying the constants of motion, and for a segment of
trajectory passing through $\bq,\bp$. Parameterize this segment with a
time-like variable $\tau$, so that, for small $\tau$,
\begin{subequations}
  \begin{align}
    \bq' &\approx \bq + \frac{\partial \bq}{\partial \tau} \tau
    \nonumber
    \\
    &= \bq + \frac{\partial H}{\partial \bp} \tau
    \intertext{and}
    \bp' &\approx \bp + \frac{\partial \bp}{\partial \tau} \tau
    \nonumber
    \\
    &= \bp - \frac{\partial H}{\partial \bq} \tau.
  \end{align}
\end{subequations}
Inserting into $K$ gives
\begin{equation}\label{eq:approxkernel1}
  K(\bq,\bp,\bq',\bp') = \exp\Bigl\{
  - \frac{\tau^2}{T^2}
  - \frac{i}{\hbar}\bp\cdot\frac{\partial H}{\partial \bp} \tau
  \Bigr\}
\end{equation}
where
\begin{equation}
  T = \biggl[
    \frac{1}{8\sigma^2}\Bigl\|\frac{\partial H}{\partial \bp}\Bigr\|^2
    + \frac{\sigma^2}{2\hbar^2}\Bigl\|\frac{\partial H}{\partial \bq}\Bigr\|^2
    \biggr]^{-1/2}.
\end{equation}
Notice that $T$ depends on $\bq,\bp$. It is non-zero except at
possible stationary points, i.e., where both $\partial H/\partial\bq$
and $\partial H/\partial\bp$ are zero.

Parameterize the dimensions other than $\tau$ with a $2D-1$
dimensional vector $s$, scaled so that $d\tau\, d^{2D-1}s =
(dq\,dp)^D$.  A local part of a wavefunction with support only on this
segment of trajectory is
\begin{equation}\label{eq:efuncdelta}
  \eta(\bq',\bp') =
  (2\pi\hbar)^D A e^{i\beta(\bq',\bp')/\hbar}
  \delta^{2D-1}(s)
\end{equation}
where $A$ is a constant. Expanding $\beta$ around $\bq,\bp$ gives
\begin{align}
  \beta(\bq',\bp') &\approx \beta(\bq,\bp)
  + \Bigl(\frac{\partial \beta}{\partial \bq} \cdot \frac{\partial \bq}{\partial \tau}
  + \frac{\partial \beta}{\partial \bp} \cdot \frac{\partial \bp}{\partial \tau}\Bigr) \tau
  \nonumber
  \\
  &= \beta(\bq,\bp)
  + \Bigl(\frac{\partial \beta}{\partial \bq} \cdot \frac{\partial H}{\partial \bp}
  - \frac{\partial \beta}{\partial \bp} \cdot \frac{\partial H}{\partial \bq}\Bigr) \tau
  \nonumber
  \\
  &= \beta(\bq,\bp) + \{\beta, H\} \tau.
\end{align}
Using Eq.~\eqref{eq:phase}, this is
\begin{equation}
  \label{eq:betatau}
  \beta(\bq',\bp') \approx \beta(\bq,\bp)
  + \Bigl(E - H + \bp\cdot\frac{\partial H}{\partial \bp}\Bigr)\tau.
\end{equation}

Inserting \eqref{eq:approxkernel1}, \eqref{eq:efuncdelta}, and
\eqref{eq:betatau} into \eqref{eq:kernelprojeta}, the projection is
\begin{multline}
  \Tilde{\eta}(\bq,\bp) \approx
  A e^{
  i\beta(\bq,\bp)/\hbar
  }
  \\
  \times \int \!d\tau
  \exp\Bigl[
  - \frac{\tau^2}{T^2}
  + \frac{i}{\hbar} (E - H)\tau
  \Bigr]
\end{multline}
\begin{equation}
  = 
  \sqrt{\pi} A T e^{i\beta(\bq,\bp)/\hbar}
  \exp\Bigl[
    -\frac{(E-H)^2T^2}{4\hbar^2}
  \Bigr].
\end{equation}
Therefore, where the original function $\eta$ had support, the
projected wavefunction $\Tilde{\eta}$ is maximal if $H$ equals the
energy eigenvalue $E$. This then is the condition for a wavefunction
like \eqref{eq:efuncdelta} to best approximate an eigenfunction in
$\hQ$. A more general approximate eigenfunction can involve manifolds
with other values of $H$, but those values must be close to $E$, where
close means
\begin{equation}
  \lvert E - H \rvert \lesssim \hbar / T.
\end{equation}
In fact as $\lvert E - H \rvert$ grows the projection approaches zero,
i.e., it converges to an element of the null space of the projection
operator.

\section{Invariance of \texorpdfstring{$p\cdot dq$}{p.dq} under coordinate transformations\label{app:pdqinv}}

Suppose that we have some old variables $q, p$ and want to transform
to new variables $Q, P$. A coordinate transformation $Q_i = f_i(q)$
can be implemented as a canonical transformation with generating
function \cite[Sec.~9-2]{goldstein}
\begin{equation}
  F(q, P) = \sum_i f_i(q) P_i.
\end{equation}
This doesn't depend explicitly on time, so the Hamiltonian does not
change values. The old momenta can be written as
\begin{equation}\label{eq:p_as_dF_dq}
  p_i = \frac{\partial F}{\partial q_i}
  = \sum_j \frac{\partial f_j}{\partial q_i} P_j.
\end{equation}
Use this and the fact that the $q$ are independent of $P$ to express
$\partial H/\partial P_i$ as a function of the $Q, P$, i.e., the
derivative of $H\bigl(q(Q, P), p(Q, P)\bigr)$, getting
\begin{align}
  \frac{\partial H}{\partial P_i} &=
  \sum_j  \Bigl[
    \frac{\partial H}{\partial q_j} \frac{\partial q_j}{\partial P_i}
    + \frac{\partial H}{\partial p_j} \frac{\partial p_j}{\partial P_i}
    \Bigr] \nonumber \\
  &= \sum_j
    \frac{\partial H}{\partial p_j}  \frac{\partial f_i}{\partial q_j}.
\end{align}
We can now write
\begin{equation}
  \sum_i P_i \frac{\partial H}{\partial P_i} =
  \sum_{i,j} P_i  \frac{\partial f_i}{\partial q_j}
  \frac{\partial H}{\partial p_j}
  = \sum_j p_j \frac{\partial H}{\partial p_j}.
\end{equation}
This and the fact that Poisson brackets are canonical invariants show
that \eqref{eq:phase2} keeps its form under this kind of
transformation.

\section{Proof that \texorpdfstring{$\{c_i, H\}_j = 0$}{\{ci,H\}j=0}\label{app:ciHj}}

If $\bq,\bp$ are separable, they must be isolated in a function within the
Hamiltonian, so it can be expressed as
$H(q_{\Bar{\imath}},p_{\Bar{\imath}}, f_i(q_i,p_i))$
\cite[Section~10-4]{goldstein}.  Here
$q_{\Bar{\imath}},p_{\Bar{\imath}}$ indicates all of the $\bq,\bp$ pairs
except the $i$-th. It is easy to show that $f_i(q_i,p_i)$ is itself a
constant of the motion, and so a function of a corresponding $c$, say
$c_i$. For a completely separable system this procedure is repeated
with successive $\bq,\bp$ pairs, each isolated in its own function within
$H$ though possibly dependent on previously separated $\bq,\bp$ pairs
through their corresponding constants of motion.

Put the indices of the $\bq,\bp$ pairs in the order of their
separation. For brevity use
\begin{equation}
  \{B, C\}_i = \frac{\partial B}{\partial q_i} \frac{\partial
    C}{\partial p_i} - \frac{\partial B}{\partial p_i}
  \frac{\partial C}{\partial q_i},
\end{equation}
so the Poisson bracket $\{B,C\}=\sum_i\{B,C\}_i$. Because they are
constants of the motion, $\Dot{c}_i =0$ for all $i$, and for $c_1$
this implies
\begin{equation}
  0 = \Dot{c}_1
  = \{c_1, H\}
  = \{c_1, H\}_1,
\end{equation}
since $c_1$ doesn't depend on $q_i, p_i$ for $i > 1$. $c_2$
depends on $q_2,p_2$ and possibly $c_1$, so
\begin{multline}
  0 = \Dot{c}_2
  = \{c_2, H\}
  = \{c_2, H\}_2 +
  \frac{\partial c_2}{\partial c_1}\{c_1, H\}_1 \\
    = \{c_2, H\}_2.
\end{multline}
Continuing this for successive $c$'s gives
\(
  0 = \{c_i, H\}_i
\)
for all $i$. We can generalize this.  For $i < j$, $c_i$ is
independent of $q_j,p_j$, so $0 = \{c_i, H\}_j$.  For $j < i$ the
expression $\{c_i, H\}_j$ does not depend on constants $c_k$ with $k <
j$, since it only has derivatives with respect to $q_j$ and $p_j$. For
$k$ with $j \leq k < i$, the constant of motion $c_i$ can depend on
$c_k$, directly or through one or more $c_l$ with $k < l < i$. This
leads to
\begin{equation}
  \{c_i, H\}_j
  = \sum_{k=j}^{i-1}\frac{\partial c_i}{\partial c_k}\{c_k, H\}_j.
\end{equation}
We can show that this is zero by induction on $i$. Therefore
Eq.~\eqref{eq:ciHj}, i.e., $\{c_i, H\}_j = 0$, holds for all $i$ and
$j$.

\bibliography{classWave.bib}

\end{document}